\documentclass[a4paper]{jpconf}
\usepackage{graphicx}

\newcommand{\be}{\begin{equation}}
\newcommand{\ee}{\end{equation}}
\newcommand{\bea}{\begin{eqnarray}}
\newcommand{\nn}{\nonumber}
\newcommand{\eea}{\end{eqnarray}}

\begin{document}
\title{Matching of analytical and numerical solutions for neutron stars of arbitrary rotation}

\author{George Pappas}

\address{Section of Astrophysics, Astronomy, and Mechanics, Department of Physics,
University of Athens, Panepistimiopolis Zografos GR15783, Athens,
Greece}

\ead{gpappas@phys.uoa.gr}

\begin{abstract}
We demonstrate the results of an attempt to match the two-soliton
analytical solution with the numerically produced solutions of the
Einstein field equations, that describe the spacetime exterior of
rotating neutron stars, for arbitrary rotation. The matching
procedure is performed by equating the first four multipole
moments of the analytical solution to the multipole moments of the
numerical one. We then argue that in order to check the
effectiveness of the matching of the analytical with the numerical
solution we should compare the metric components, the radius of
the innermost stable circular orbit ($R_{ISCO}$), the rotation
frequency $\Omega\equiv\frac{d\phi}{dt}$ and the epicyclic
frequencies $\Omega_{\rho},\;\Omega_z$. Finally we present some
results of the comparison.
\end{abstract}

\section{Introduction}
\label{sec:1}

Obtaining the geometry of the spacetime that surrounds a rotating
distribution of mass in an exact form, is a problem of
astrophysical relevance that remains unsolved. When modelling the
astrophysical phenomena that we observe around rotating neutron
stars, there are only a few choices we can make for the geometry
we will use. We can either use one of the very well understood
geometries of Schwarzschild or Kerr, which can only be used as
very rough approximations, or we have to use a numerically
obtained spacetime.

Although there are various groups (see \cite{Sterg}, and for an
extended list of numerical schemes see \cite{Lrr}), that have
expertise in building relativistic models of astrophysical objects
with adjustable  physical characteristics and can construct the
metric inside and outside such objects by solving numerically the
full Einstein equations in stationary cases, having the metric of
the spacetime in tabulated form with numerical values that
correspond to the metric components at a selected grid is not
always appropriate for studying the phenomena around the object.

Thus, an analytical solution that could describe the geometry
around such an object would be useful. Fortunately, nowadays,
there is a large variety of analytical solutions of vacuum
Einstein equations, which could be used as candidate metrics to
describe well the exterior space-time of axisymmetric
astrophysical objects. Ernst \cite{ernst1} formulated the Einstein
equations in the case of axisymmetric stationary space-times long
time ago, while Manko et al. and Sibgatullin
\cite{twosoliton,manko,manko2,manko3,sib1,SibManko} have used
various analytical methods to produce such spacetimes
parameterized by various parameters that have a different physical
context depending on the type of each solution.

Such an analytical solution, appropriately matched to a neutron
star, could be used to describe the stationary properties of the
spacetime. That is, study the geodesics in the exterior of the
neutron star. In other words, from the analytical solution, we
could obtain bounds of motion for test particles, the location of
the innermost stable circular orbit (ISCO), the rotation frequency
of the circular orbits on the equatorial plane as well as the
epicyclic frequencies around them.

These properties that are of astrophysical relevance and
characteristic of a spacetime, will be used to compare the
proposed analytical solution to the numerically produced solutions
that describe the spacetime exterior to a neutron star.

The rest of the paper is organized as follows: In Sec.~\ref{sec:2}
the proposed analytical solution (two-soliton) is briefly
presented and some of its properties analyzed. In Sec.~\ref{sec:3}
we present at first the matching criteria and then the comparison
criteria and some results.

\section{An Analytical axially symmetric solution: The Two Soliton solution.}
\label{sec:2}

The vacuum region of a stationary and axially symmetric spacetime
can be described by the Papapetrou line element \cite{Pap}

\be \label{Pap} ds^2=-f\left(dt-\omega d\phi\right)^2+
      f^{-1}\left[ e^{2\gamma} \left( d\rho^2+dz^2 \right)+
      \rho^2 d\phi^2 \right],
\ee

\noindent were the functions $f,\;\omega$ and $\gamma$ are
functions of the Weyl-Papapetrou coordinates ($\rho,\;z$). Using
this metric, the Einstein field equations in vacuum reduce to the
Ernst equation \cite{ernst1}

\be \label{ErnstE}
Re(\mathcal{E})\nabla^2\mathcal{E}=\nabla\mathcal{E}\cdot\nabla\mathcal{E},
\ee

\noindent were the Ernst potential $\mathcal{E},$ is a complex
function of the metric functions.

A general procedure for generating solutions of the Ernst equation
was developed by Sibgatullin and Manko \cite{sib1,SibManko,manko2,
twosoliton}. The solution of the Ernst equation is produced from a
choice of the Ernst potential on the axis of the form

\be \mathcal{E}(\rho=0,z)=e(z)=\frac{P(z)}{R(z)}, \ee

\noindent were the functions $P(z), R(z)$ are polynomials of order
$n$ with complex coefficients in general. In \cite{manko2}, one
can find a detailed description of the procedure for generating
the solutions using the parameters of the Ernst potential. It is
shown in \cite{manko2} that the metric functions are finally
expressed in terms of some determinants, which depend on the
parameters $\xi_n$, which are the $2n$ roots of the equation
$e(z)+e(z)^*=0$, the parameters $e_n,\;f_n$ and $\beta_n$, which
are defined from the expressions
\[e(z)=1+\sum^n_{k=1}\frac{e_k}{z-\beta_k},\;\; f(z)=\sum^n_{k=1}\frac{f_k}{z-\beta_k}\]
and finally the expressions $R_n=\sqrt{\rho^2+(z-\xi_n)^2}$, which are functions of the coordinates $(\rho,z)$.\\
\\
\indent The vacuum two-soliton solution (proposed by Manko
\cite{twosoliton}) is a special case of the previous general
axisymmetric solution that is obtained from the ansatz (see also
\cite{PapSotiri})

\be \label{2soliton}
e(z)=\frac{(z-M-ia)(z+ib)-k}{(z+M-ia)(z+ib)-k}\ee

\noindent where all the parameters are real while the parameters
$M,\;a$ are the mass and the reduced angular momentum
$\frac{J}{M}$ respectively. The first five mass and angular
momentum moments of the corresponding spacetime are:

\bea \label{moments2}
M_0&=&M,\quad M_1=0,\quad M_2=-(a^2-k)M,\quad M_3=0,\nn\\
 M_4&=&\left( a^4 - (3a^2 -2ab + b^2)k + k^2
+\frac{1}{7}(2a^2-k)M^2\right) M\nn\\
J_0&=&0,\quad J_1=aM,\quad J_2=0,\nn\\
J_3&=&- (a^3 -(2a - b)k)M  ,\quad J_4=0. \eea

\noindent For the ansatz (\ref{2soliton}), the equation
$e(z)+e(z)^*=0$ corresponds to:

\be \label{poly} z^4-(M^2-a^2-b^2+2k)z^2+(k-ab)^2-b^2M^2=0,\ee

\noindent where the coefficients take real values. That means that
the roots can either be real, or conjugate pairs. We can also see
that the four roots of (\ref{poly}) are of the form
\[\xi_1=-\xi_3=\xi_+,\;\;\xi_2=-\xi_4=\xi_-.\]

\noindent Defining the parameters $d=\sqrt{(k-ab)^2-b^2M^2}$ and
$\kappa_{\pm}=\sqrt{M^2-a^2-b^2+2k\pm2d}$, we can express the
roots of (\ref{poly}) as

\be\xi_{\pm}=\frac{1}{2}\left(\kappa_+\pm\kappa_-\right).\ee

\noindent Using these parameters we redefine $R_n$ as

\be R_{\pm}=\sqrt{\rho^2+(z\pm
\xi_+)^2},\;r_{\pm}=\sqrt{\rho^2+(z\pm \xi_-)^2}.\ee

Depending on the choice of the parameters used in
(\ref{2soliton}), there can be obtained various types of
solutions. For example, if one sets the value of $k$ equal to
zero, for arbitrary values of the parameter $b$, the two-soliton
corresponds to the Kerr metric of mass $M$ and reduced angular
momentum $a$ (if $|a|<M$) or the hyperextreme Kerr (if $|a|>M$).
Further more, if one chooses $a$ equal to zero, then the
two-soliton corresponds to the Schwarzschild solution of mass $M$.

There is also another interesting special case for the two-soliton
solution. If the parameter $d$ is real and we have the conditions
$M^2-a^2-b^2+2k+2d>0$ and $M^2-a^2-b^2+2k-2d=0$, then the
parameter $\kappa_+$ is real and the parameter $\kappa_-=0$. In
that case there is a degeneracy and there are only two real roots
$\xi_{1,2}=\pm \xi_{\pm}=\pm\frac{\kappa_+}{2}$. This solution
corresponds to the Manko et al. solution \cite{manko,manko3}.

In brief, the two-soliton solution can produce a very rich family
of analytical solutions including the classical solutions of
Schwarzschild and Kerr as well as the Manko et al. solution
\cite{manko,manko3} that has been used in \cite{BertiSterg,Stuart}
to approximately match the exterior of a rotating neutron star.

At this point we can point out the advantages of the two-soliton
solution against the Manko et al. solution \cite{manko,manko3}. If
we assume a constant mass $M$, the two-soliton solution has a
three dimensional parameter space $(a,b,k)$. In that space the
plane $k=0$ includes all Kerr solutions of mass $M$ while the $b$
axis corresponds to the Schwarzschild solution of mass $M$.

\begin{figure}

\includegraphics[width=.33\textwidth]{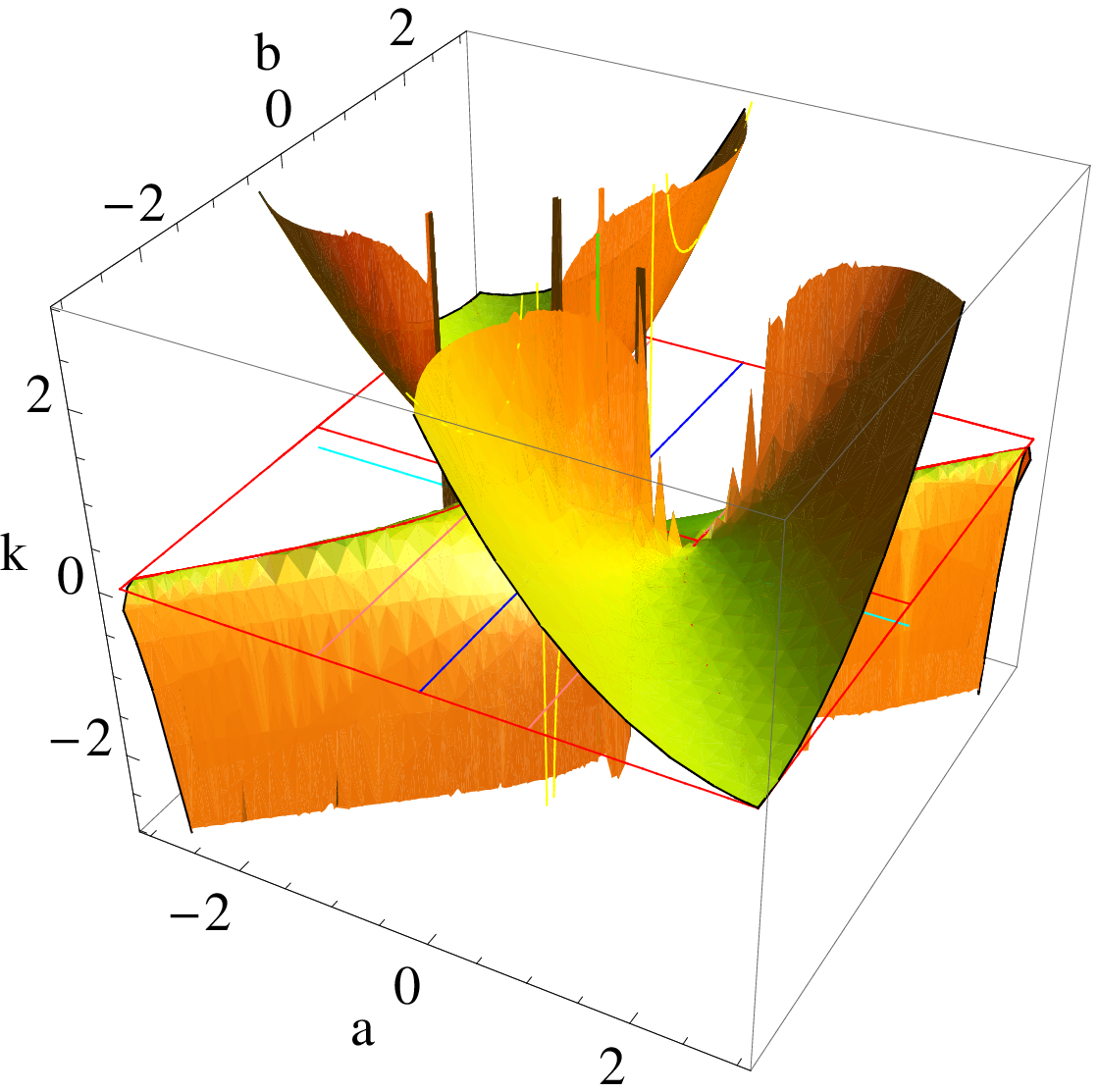}
\includegraphics[width=.33\textwidth]{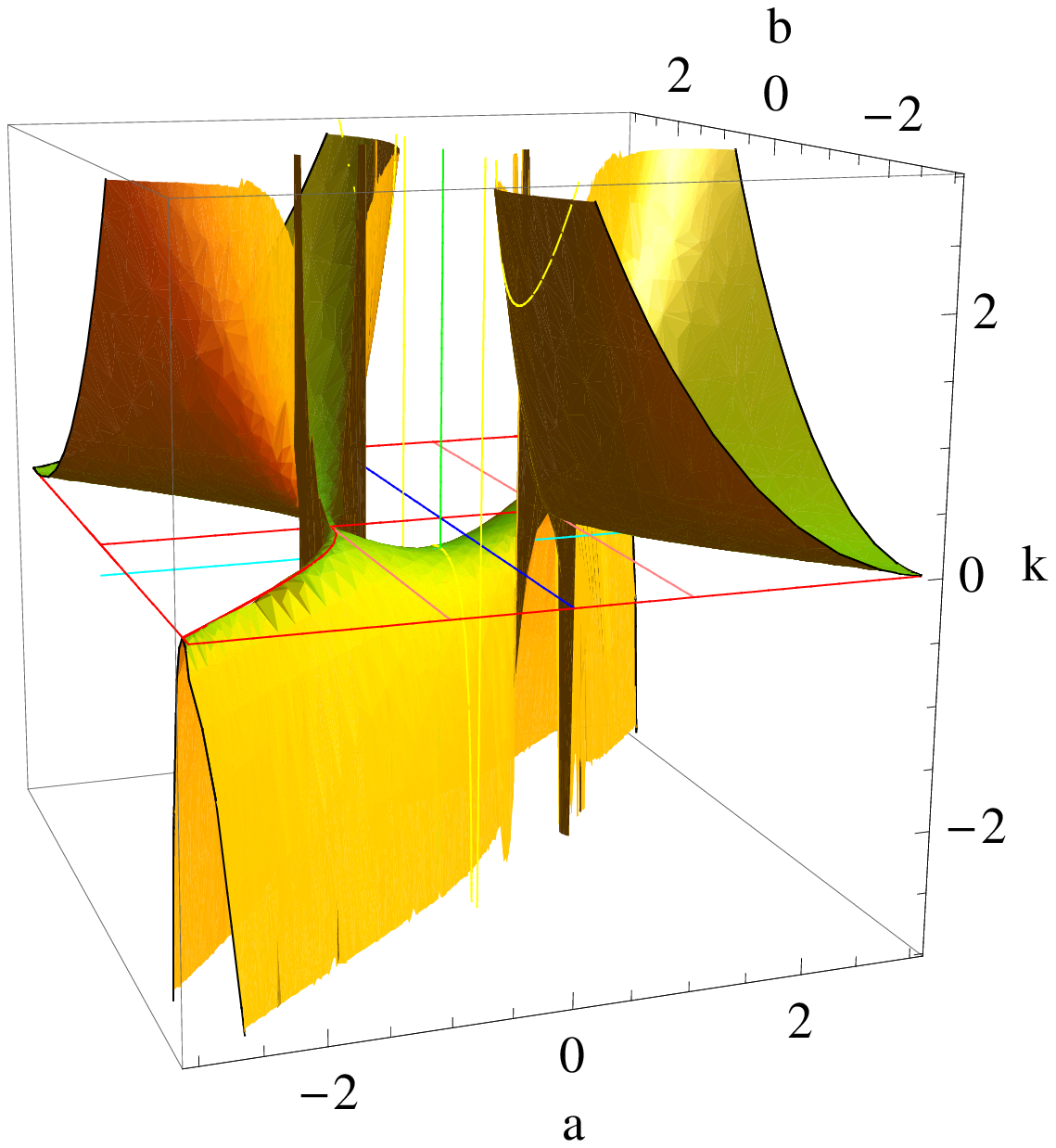}
\includegraphics[width=.33\textwidth]{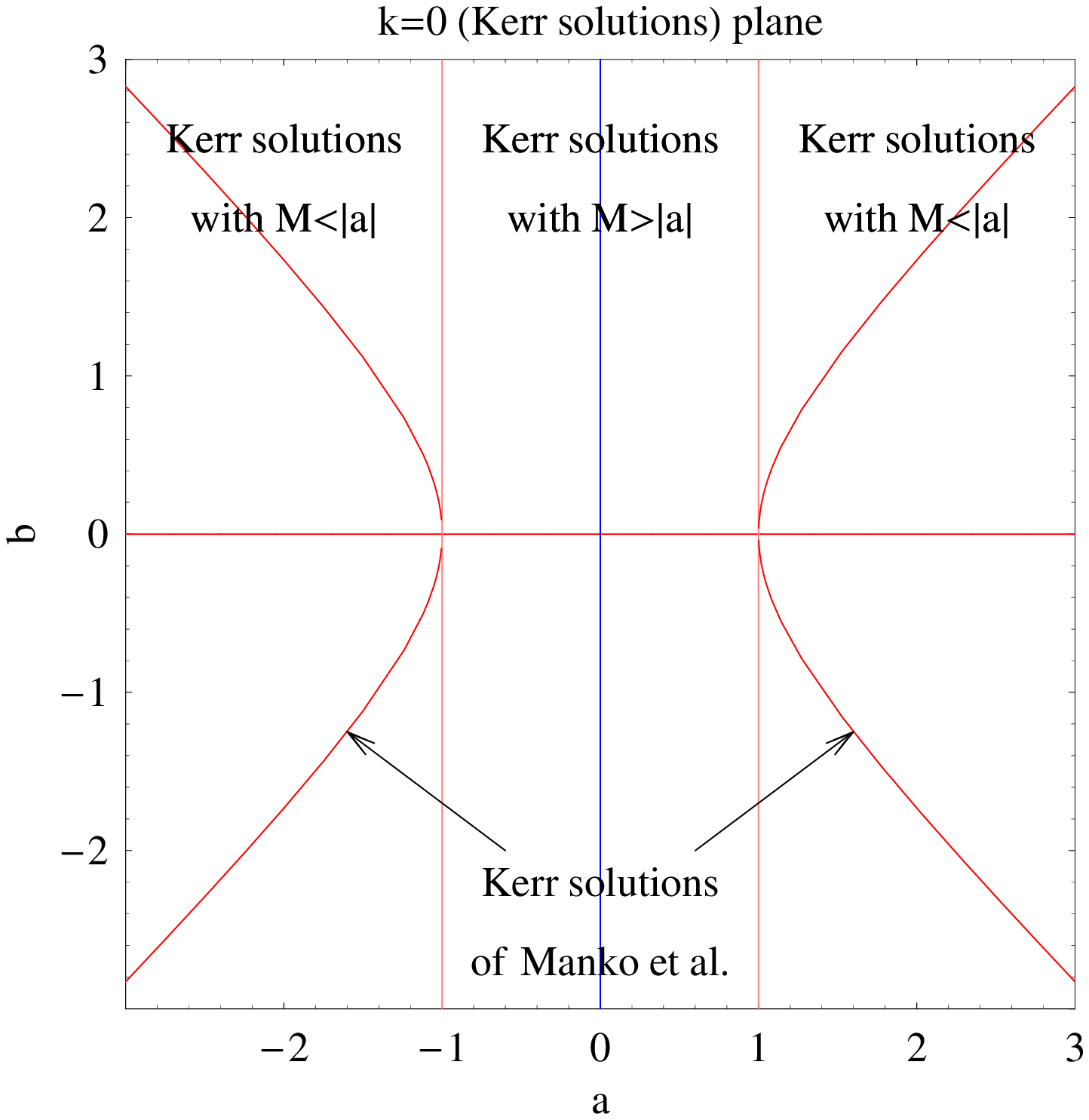}
\caption{The first two plots show the parameter space $(a,b,k)$ of
the two-soliton solution for a specific mass ($M=1$) from two
different view points. The 2D surface plotted is the constrain of
$k$ which corresponds to the Manko et al. solution
\cite{manko,manko3} used by Berti and Stergioulas
\cite{BertiSterg}. The third plot is the $k=0$ plane of the
parameter space which corresponds to all Kerr solutions of the
two-soliton solution. The plots clearly show that the Manko et al.
solution has no set of parameters that describe the case of $k=0$
which corresponds to the Kerr and the Schwarzschild (for $a=0$)
solutions, since there is no intersection between the constrain of
$k$ and the Kerr plane within the appropriate range of parameters.
The two hyperbola plotted on the plane $k=0$ are the points were
the constrain of $k$ touches the plane in a tangent manner. As we
can see these hyperbola correspond to $|a|>M$.}
\protect\label{figure1}

\end{figure}

As we have mentioned earlier, the Manko et al. solution
\cite{manko,manko3} is obtained if we impose the constrain
$M^2-a^2-b^2+2k-2d=0$ to the parameters. This is analogous to
having a two dimensional surface in the parameter space. There are
two things that could be pointed out about that surface, which is
drawn in Fig.~\ref{figure1}. The first is that the axis of $b$
(the blue line) that corresponds to the Schwarzschild solution
doesn't cross the surface and the second is that the Kerr
solutions with $M>|a|$, that lie in the parameter space on the
plane $k=0$ between the two red lines parallel to the $b$ axis, do
not intersect the surface either.

Thus, this solution can reduce to Schwarzschild or Kerr only
formally. These are important drawbacks and were pointed out by
Berti and Stergioulas \cite{BertiSterg}. Particularly important is
the fact that it can only be matched with rapidly rotating models
of neutron stars, since in the limit of zero rotation it has a
non-vanishing quadrupole moment
$Q=-\frac{M}{4}\frac{(M^2+b^2)^2}{(M^2-b^2)}$. These problems do
not exist for the two-soliton solution, that can represent neutron
stars of arbitrary rotation. Another important aspect of the
two-soliton is that, as can be seen from the multipole moments
(\ref{moments2}), the solution can describe spacetimes with an
arbitrary quadrupole $M_2$ and an arbitrary current octupole
$J_3$, since every parameter is introduced linearly in the first
moment it appears and thus it can always be determined by the
first four non-zero moments.

\section{Matching and comparison of the analytical to the numerical metric.}
\label{sec:3}

\subsection{ Matching of the analytical to the numerical
solution.}

In matching the analytical solution to the numerical one it is
desirable to find a criterion that will be characteristic of the
whole structure of the spacetime and not of a small region of it.
That is, the matching conditions should be global and not local.
Berti and Stergioulas \cite{BertiSterg} have argued that these
global conditions should be the matching of the multipole moments.
We will work under the same assumption. The multipole moments of a
stationary axially symmetric spacetime are connected to the Ernst
potential on the axis of symmetry and the spectrum of the moments
fully specify it. On the other hand, when the Ernst potential on
the axis is given, there exists a spacetime which is fully
specified by that Ernst potential \cite{Xanth}. Thus, the
multipole moments are characteristic of a spacetime and can be
used as a global matching condition.

When a numerical spacetime for a neutron star model is constructed
from a numerical algorithm, it is possible to evaluate its mass
moments $M,\,Q,...$ and angular momentum moments $J,\,S_3,...$
(for further discussion see Berti and Stergioulas
\cite{BertiSterg}). These numerically evaluated moments can be
used to impose the matching conditions to the analytical
spacetime. The first four nonzero multipole moments of the
two-soliton solution are:

\bea \label{moments3} M_0=M&,& J_1=aM,\nn\\
M_2=-(a^2-k)M&,&J_3=- (a^3 -(2a - b)k)M, \eea

\noindent where we can see that once we specify the mass and the
angular momentum of the spacetime, the parameter $k$ is uniquely
determined by the quadrupole moment $Q=M_2$ and the parameter $b$
is uniquely determined by the current octupole $S_3=J_3$. Thus,
having specified the parameters of the two-soliton, we have
completely determined the analytical spacetime that can be used to
describe the exterior of the neutron star. What remains to be seen
is how well do the properties of the analytical spacetime compare
to those of the numerical one.\\

\subsection{Criteria for the comparison of the analytical to the numerical spacetime.}

\indent To compare the analytical to the numerical spacetime, one
should use criteria that are characteristic of the geometric
structure of the spacetime. The criteria should also be related to
properties of the spacetime that could potentially be measured in
astrophysical phenomena. A good fit of such properties would mean
that they could be used to identify a spacetime through
observational data and then associate it with properties of the
compact object that generates it.

The first and more obvious criterion one can use is how well the
respective components of the analytical and the numerical metric
match. If the relative difference of the analytical metric
components

\be g_{tt}=-f,\quad g_{t\phi}=f\omega,\quad
g_{\phi\phi}=f^{-1}\rho^2-f\omega^2 \ee

\noindent to the corresponding numerical is small, one could
consider the former as a good substitute of the latter. The
comparison of the $g_{\phi\phi}$ component can be performed
through the circumferential radius $R_{circ}=\sqrt{g_{\phi\phi}}$
which is as an indicator of how well the measurements of the
circumferences in the corresponding spaces (analytical and
numerical) with a specific value of the coordinate $\rho$, compare
with each other.

The second criterion is the innermost stable circular orbit
(ISCO). Test particles that move on geodesics on the equator
follow trajectories that are given by the equation

\be \label{Veff}
-g_{\rho\rho}\left(\frac{d\rho}{d\tau}\right)^2=1-
\frac{\tilde{E}^2g_{\phi\phi}+2\tilde{E}\tilde{L}g_{t\phi}+\tilde{L}^2g_{tt}}{\rho^2}\equiv
V(\rho) \ee

\noindent where $\tilde{E}$ and $\tilde{L}$ are the conserved
energy per unit mass and angular momentum (parallel to the axis
component) per unit mass and $V(\rho)$ is the effective potential.
Since we are interested in circular orbits, the conditions for
such orbits are $d\rho/d\tau=0$ and $d^2\rho/d\tau^2=0$ which are
equivalent from (\ref{Veff}) to the conditions for the effective
potential, $V(\rho)=0,\quad dV(\rho)/d\rho=0$. The ISCO is
determined if we additionally impose the condition that the radius
$\rho$ of the circular orbit is also a turning point of the
potential, that is $d^2V(\rho)/d\rho^2=0$. The location of the
ISCO is of astrophysical importance since it indicates the radius
at which matter orbiting a compact object can no longer maintain
its orbit and plunges into the object. That radius would be the
internal radius of accretion disks formed around such a compact
object.

The third criterion is the rotation frequency of circular orbits
on the equatorial plane. The rotation frequency
$\Omega\equiv\frac{d\phi}{dt}$ is determined from (\ref{Veff}) and
the conditions for circular orbits, $V(\rho)=0,\quad
dV(\rho)/d\rho=0$, and is given by the equation

\be
\Omega(\rho)=\frac{-g_{t\phi,\rho}+\sqrt{(g_{t\phi,\rho})^2-g_{tt,\rho}g_{\phi\phi,\rho}}}{g_{\phi\phi,\rho}}.
\ee

The final criterion are the epicyclic frequencies
$\Omega_{\rho},\;\Omega_z$. These are the frequencies of
precession of the periastron and the orbital plane of a perturbed
circular orbit, respectively. They can be evaluated by perturbing
the equations of motion around a circular orbit

\be
-g_{\rho\rho}\left(\frac{d\rho}{d\tau}\right)^2-g_{zz}\left(\frac{dz}{d\tau}\right)^2=
V(\rho,z) \ee

\noindent where $V(\rho,z)$ is the effective potential defined in
(\ref{Veff}) maintaining its $z$ dependence. The final expressions
for the epicyclic frequencies are

\bea \Omega_a&=&\Omega-\left\{-\frac{g^{aa}}{2}\left[(g_{tt}+g_{t\phi}\Omega)^2\left(\frac{g_{\phi\phi}}{\rho^2}\right)_{,aa}\right.\right.\nn\\
              &&-2(g_{tt}+g_{t\phi}\Omega)(g_{t\phi}+g_{\phi\phi}\Omega)\left(\frac{g_{t\phi}}{\rho^2}\right)_{,aa}\nn\\
              &&\left.\left.(g_{t\phi}+g_{\phi\phi}\Omega)^2\left(\frac{g_{tt}}{\rho^2}\right)_{,aa}\right]\right\}^{1/2},
\eea

\noindent where $a$ is $\rho$ or $z$ and the expression is
evaluated at $z=0$ and around circular orbits of $\rho=\rho_0$.

These criteria can be related to astrophysical phenomena and in
particular, they can be related to properties and observables of
accretion disks. One such observable is the quasi-periodic
oscillations (QPOs) with modulations in the range of kHz that have
been observed in the X-ray flux from accreting X-ray pulsars
\cite{StergQPO}. The association of QPOs with the above
frequencies could be used to determine the properties of the
central object from the values of the various frequencies. These
properties such as mass, angular momentum and mass quadrupole,
could also be used to gain insight on the structure of the central
object. Thus, an analytical spacetime that fits well all these
parameters could be practically considered equivalent to the
numerical one and a useful tool for doing astrophysics.\\

\subsection{Results of the comparison.}

\indent In the comparison of the metric components, we have used
the two corresponding Manko et al. solutions that were used by
Berti and Stergioulas \cite{BertiSterg}, as a reference. The
comparison of the metrics is performed on the equatorial plane and
on the axis of rotation. Here we will present only one rotating
neutron star model, since these results are typical for all
models. The chosen example is the one in table 3 of the Berti and
Stergioulas \cite{BertiSterg} work, that corresponds to the model
with the fastest rotation of the sequence of maximum mass in the
non-rotating limit, for the FPS equations of state (EOS).

\begin{figure}[h]

\begin{minipage}{18pc}
\includegraphics[width=1\textwidth]{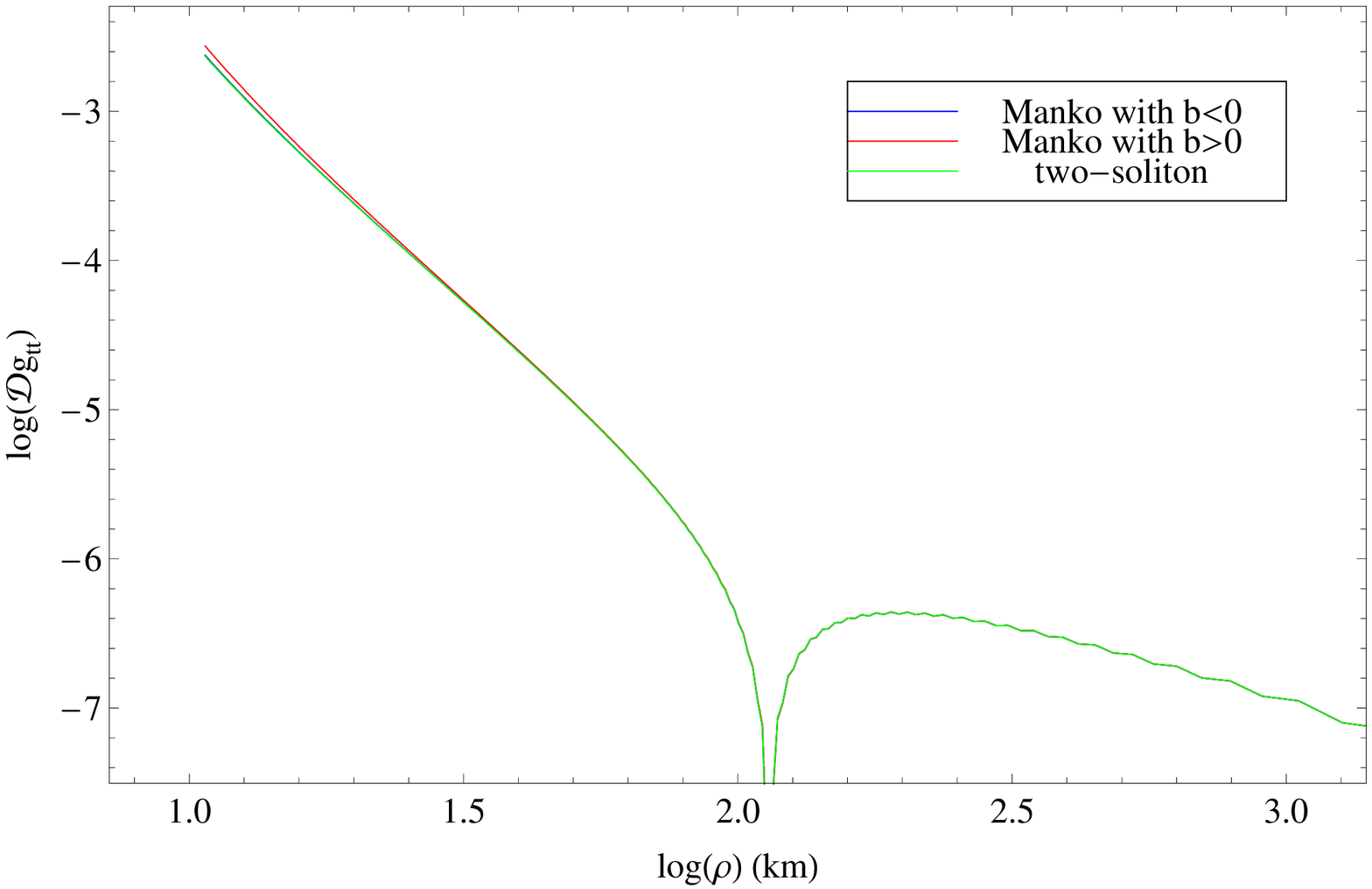}
\includegraphics[width=1\textwidth]{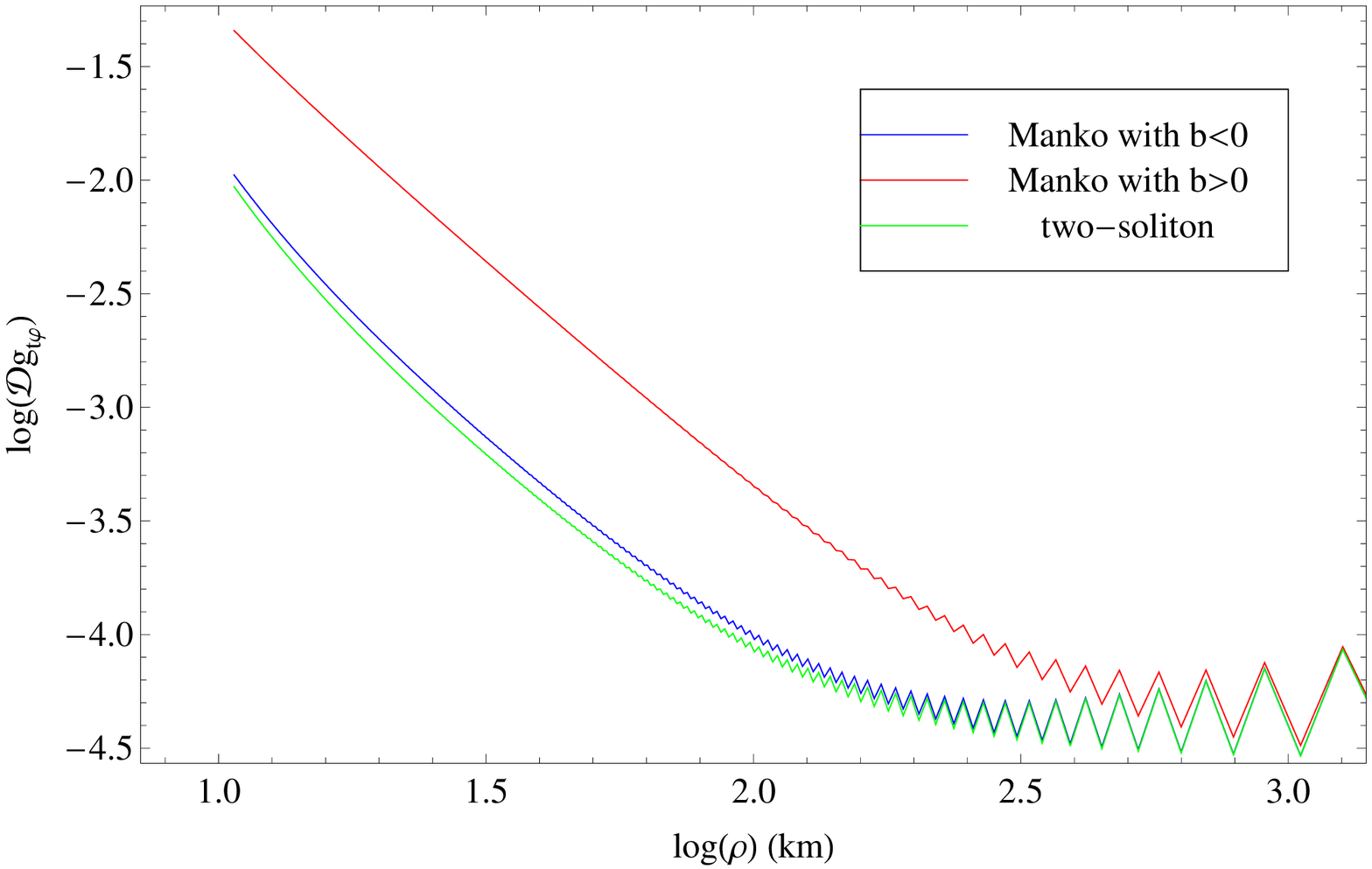}
\includegraphics[width=1\textwidth]{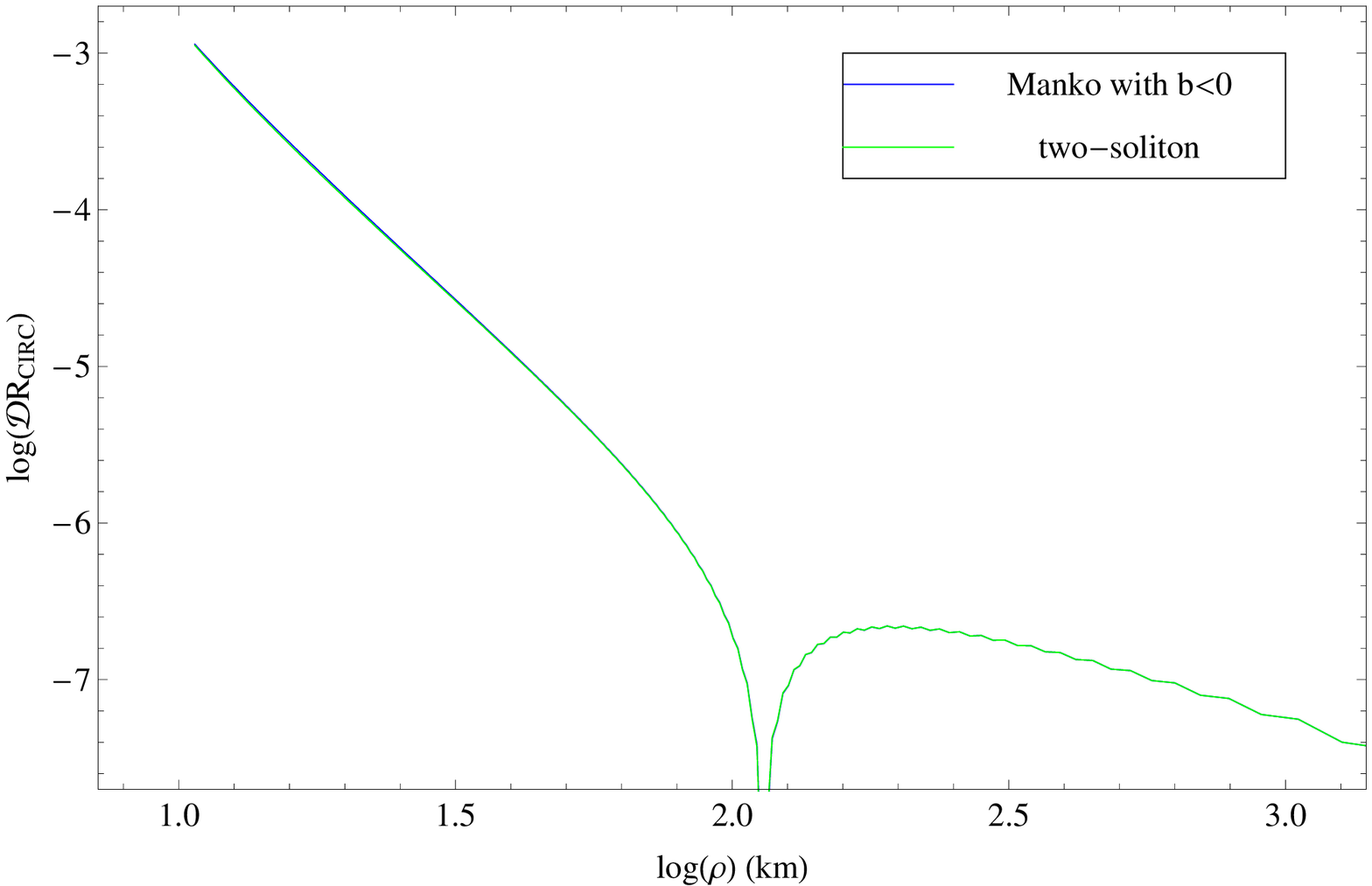}
\end{minipage}\hspace{1.5pc}%
\begin{minipage}{18pc}

\caption{The first two plots show the relative difference of the
three analytical metrics with the numerical for the components
$g_{tt}$ and $g_{t\phi}$ on the equatorial plane. The analytical
metrics used are the two-soliton and the two Manko et al.
\cite{manko,manko3} that correspond to the two values of the
parameter $b$ (see Berti and Stergioulas \cite{BertiSterg}). The
third plot shows the relative difference of the circumferential
radius on the equatorial plane for the two-soliton and the Manko
et al. with the negative value of $b$ (the two solutions almost
coincide). As we can see, the errors are well under 1 percent (for
the green line) on the surface and fall outwards. The oscillations
that appear in the second plot are due to the behavior of the
numerical solution.} \protect\label{figure2}
\end{minipage}

\end{figure}

In Fig. \ref{figure2} we have plotted the relative difference of
the metric components $g_{tt}$ and $g_{t\phi}$ and of the
circumferential radius $R_{circ}$ as a function of coordinate
radius $\rho$ on the equatorial plane (the $\mathcal{D}$ operator
is defined as $\mathcal{D}=$(numerical - analytical)/numerical).
In the first two graphs we have used for comparison the Two
Soliton solution and the two corresponding solutions of Manko et
al. which are distinguished by their value of the parameter $b$ of
the Ernst potential. We remind here that the Ernst potential of
the Manko et al. solution is given by the expression

\bea
e(z)=\frac{(z-M-ia)(z+ib)+d-\delta-ab}{(z+M-ia)(z+ib)+d-\delta-ab},\\
\delta=\frac{-M^2b^2}{M^2-(a-b)^2},\quad
d=\frac{1}{4}\left(M^2-(a-b)^2\right)\nn \eea

\noindent where there are only three independent parameters
$M,\;a,\;b$ and for a particular quadrupole moment of the
numerical solution (with fixed $M,\;a$) there may be no
corresponding value of $b$ or there may be two values. For these
two values of $b$ we can calculate two metrics which are the ones
shown in the graphs. As the second graph shows, the one of the two
Manko et al. solutions does not fit well with the numerical metric
and thus it describes it poorly \cite{BertiSterg}. On the other
hand, the other solution (corresponding to the negative value of
$b$) is as good as the two-soliton metric. Both metrics are better
than 1 percent on the equator and the relative difference falls
with increasing radial coordinate $\rho$. The relative difference
shows a kink at about 100km which is an artifact due to a crossing
of the numerical and the analytical solutions and has no further
physical meaning. In the third graph of Fig. \ref{figure2} we have
used only the one of the Manko et al. metrics (the one with
negative $b$) and the two-soliton. The relative difference for the
$R_{circ}$ is under 0.1 percent on the surface and falls with
increasing radial coordinate $\rho$.

\begin{figure}[h]

\includegraphics[width=18pc]{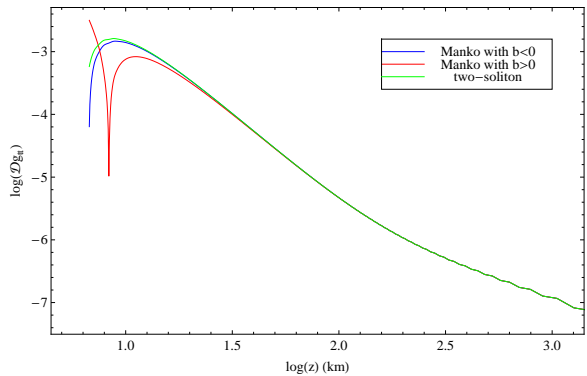}\hspace{1.5pc}%
\begin{minipage}[b]{18pc}\caption{\label{figure3}The plot shows the relative difference of the three
analytical metrics with the numerical one for the component
$g_{tt}$ on the axis of symmetry. The relative difference is under
1 percent on the surface. The behavior is similar to the behavior
on the equatorial plane.}
\end{minipage}

\end{figure}

In Fig. \ref{figure3} we have plotted the relative difference of
$g_{tt}$ on the axis of symmetry. The plot shows the same good
behavior of the analytical metric on the axis as on the equatorial
plane. The behavior on the axis of the analytical metric was a
matter of confusion that has been clarified in \cite{Pappas}.

As we can see, both the Manko et al. and the two-soliton fit well
the numerical metric. But for the Manko et al. there are
constrains. If the rotation rate of the neutron star becomes
smaller than a critical value (for typical values of the rotation
parameter $j=J/M^2$ see \cite{BertiSterg}), there cannot be found
any solutions (based on the first four multipole moments) of the
type of Manko et al. that could match to the exterior of the
neutron star. On the other hand, the two-soliton can describe
arbitrary rotations as was pointed out and the good fit to the
numerical metric extends to all rotations.

\begin{figure}[h]

\includegraphics[width=18pc]{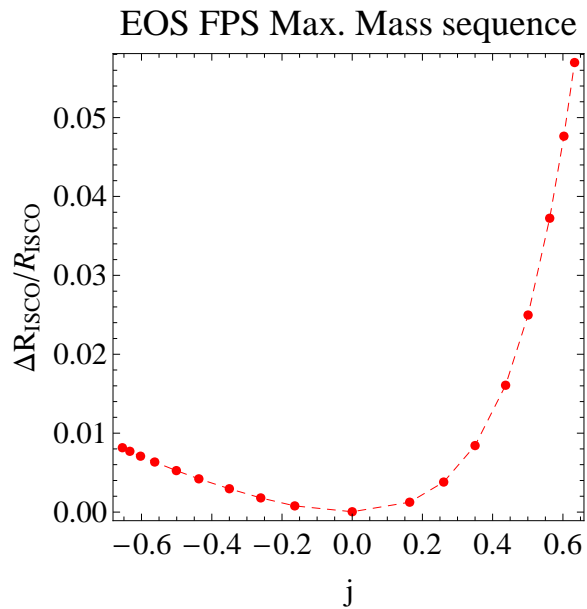}\hspace{1.5pc}
\begin{minipage}[b]{18pc}
\caption{\label{figure4}Relative difference of the analytical with
the numerical ISCO for the sequence of maximum mass in the
non-rotating limit for the equation of state FPS. The maximum
difference of the co-rotating ISCO is 5.7 percent for the fastest
rotating model that has a co-rotating ISCO and the maximum
difference for the counter rotating ISCO is less than 1 percent
for the fastest rotating model.}
\end{minipage}

\end{figure}

For the comparison of the ISCO we have used a sequence of models
of neutron stars with various rotations. The sequence is the one
in table 3 of Berti and Stergioulas \cite{BertiSterg} that
corresponds to the model of maximum mass in the non-rotating limit
for the FPS equation of state. The comparison of the $R_{ISCO}$ is
shown in Fig. \ref{figure4}, where we have plotted the relative
difference of the numerical ISCO with the analytical one as a
function of the rotation parameter $j\equiv J/M^2$. The negative
values of $j$ correspond to the counter-rotating ISCOs which are
located further from the star surface. In some cases the
co-rotating ISCO is under the surface of the star so, there is no
numerical value for it, while that never happens for the
counter-rotating ISCO. The co-rotating analytical ISCO is
systematically less accurate than the counter-rotating one and
that is to be expected since the latter is located further from
the star. The difference in the ISCOs is less than 5.7 percent for
the co-rotating and less than 1 percent for the counter-rotating
orbits. The same general results are true for the Manko et al.
metric for rotation parameters above the critical value. We should
point out that the results of Fig. \ref{figure4} are different
than the ones shown in Fig. 8 of \cite{BertiSterg}. The difference
is on the counter-rotating branch and it is due to a calculation
slip that resulted to overestimation of the error in the
counter-rotating ISCOs in \cite{BertiSterg}. For example the
fastest rotating model of fig. 7 in \cite{BertiSterg} is given an
error of 10 percent while the true error is of the order of 1.2
percent. Thus the ISCO comparison is more favorable than it was
originally estimated.

\begin{figure}[h]

\includegraphics[width=18pc]{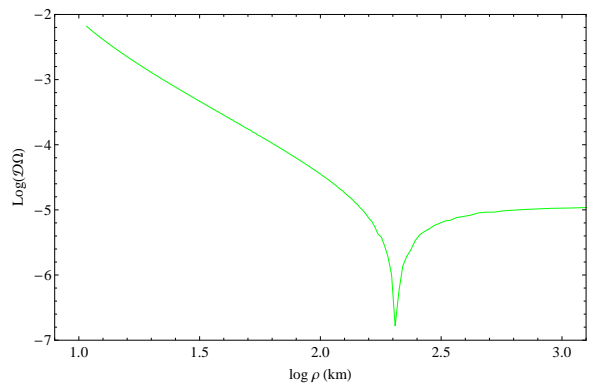}\hspace{1.5pc}
\begin{minipage}[b]{18pc}
\caption{\label{figure5}Relative difference in $\Omega$ of the
two-soliton with the numerical metric. As in the plots of the
various metric components, the relative difference is under 1
percent on the surface and falls outwards.}
\end{minipage}
\end{figure}

The next property that we compare is the rotation frequency of the
circular orbits. The relative difference of $\Omega$ on the
equatorial plane is plotted as a function of the coordinate
distance $\rho$ and is shown in Fig. \ref{figure5}. The relative
difference is less than 1 percent on the surface of the star and
reduces with increasing distance $\rho$. In this model, since the
ISCO is located under the surface of the neutron star, there is no
problem to evaluate $\Omega$ from the surface outwards. The
analytically evaluated $\Omega$, as Fig. \ref{figure5} shows, fits
well the numerical $\Omega$ and that conclusion also holds for all
rotations.

\subsection{Conclusions}

That was a brief presentation of our results on comparing the
analytical two-soliton metric with numerical metrics describing
the exterior of neutron stars. We plan to present a more thorough
comparison between the two kinds of metrics in a forthcoming paper
\cite{inprep} where the comparison of the epicyclic frequencies
will be demonstrated as well.

The results presented here suggest that the two-soliton metric is
a very good candidate for describing the exterior spacetime of a
neutron star of arbitrary rotation. Therefore, we could use this
metric to gain information about the neutron star through
observables related to astrophysical processes in the vicinity of
the star.

\ack I would like to thank Theocharis Apostolatos and Nikolaos
Stergioulas for many useful discussions. This work was supported
by the research funding program ``Kapodistrias'' with Grant No
70/4/7672.


\section*{References}
\begin{thebibliography}{9}

\bibitem{Sterg} Stergioulas~N and Friedman~J~L 1995 {\it ApJ } {\bf 444} 306.

\bibitem{Lrr} Stergioulas~N 2003 Rotating Stars in Relativity \textit{Living Rev. Relativity} {\bf 6} 3.

\bibitem{ernst1} Ernst~F~J  1968 {\it Phys. Rev.} {\bf 167} 1175,  Ernst~F~J  1968, {\it Phys. Rev.} {\bf 168} 1415.

\bibitem{twosoliton} Manko~V~S, Martin~J and Ruiz~J~E 1995 {\it J. Math. Phys.} {\bf
36} 3063.

\bibitem{manko} Manko~V~S, Sanabria-G\'omez~J~D and Manko~O~V  2000 {\it Phys. Rev.} D {\bf 62}
044048.

\bibitem{manko2} Ruiz~E, Manko~V~S and Martin~J 1995 {\it Phys. Rev.} D {\bf 51}
4192.

\bibitem{manko3} Manko~V~S, Mielke~E~W and Sanabria-G\'omez~J~D 2000 {\it Phys. Rev.} D {\bf
61} 081501.

\bibitem{sib1} Sibgatullin~N~R 1984 {\it Oscilations and Waves in Strong
Gravitational and Electromagnetic Fields} (Nauka, Moscow, 1984;
English translation: Springer-Verlag, Berlin, 1991).

\bibitem{SibManko} Manko~V~S and Sibgatullin~N~R 1993 {\it Class. Quantum
Grav.} {\bf 10} 1383.

\bibitem{Pap} Papapetrou~A 1953 {\it Ann.Phys.} {\bf 12} 309.


\bibitem{PapSotiri}
  Sotiriou~T~P and Pappas~G 2005
  {\it J.\ Phys.: Conf.\ Ser.\ }  {\bf 8} 23
  ({\it Preprint} gr-qc/0504122).

\bibitem{BertiSterg} Berti~E and Stergioulas~N 2004 {\it MNRAS }, {\bf 350}
1416.

\bibitem{Stuart} Stute~M and Camenzind~M 2002 {\it MNRAS } {\bf 336} 831.

\bibitem{Xanth} Hoenselaers~C, Kinnersley~W and
Xanthopoulos~B~C 1979 {\it Phys. Rev. Lett. } {\bf 42} 481,
Hoenselaers~C, Kinnersley~W and Xanthopoulos~B~C 1979 {\it J.
Math. Phys. } {\bf 20} 2530, Xanthopoulos~B~C 1979 {\it J. Phys.
A: Math. Gen. } {\bf 12} 1025, Xanthopoulos~B~C 1981 {\it J. Math.
Phys. } {\bf 22} 1254, Hauser~I and Ernst~F~J 1981 {\it J. Math.
Phys. } {\bf 22} 1051.

\bibitem{StergQPO}
  Kluzniak~W, Abramowicz~M~A, Kato~S, Lee~W~H and
  Stergioulas~N 2004
  {\it Astrophys.\ J.\ }  {\bf 603} L89
  ({\it Preprint} astro-ph/0308035).


\bibitem{Pappas}
  Pappas~G and Apostolatos~T~A 2008
  {\it Class.\ Quant.\ Grav.\ }  {\bf 25} 228002
  ({\it Preprint} 0803.0602 [gr-qc]).

\bibitem{inprep} Pappas~G and Apostolatos~T~A, {\it Work in
preperation}.

\end{thebibliography}
\end{document}